\documentclass[twocolumn,showpacs,preprintnumbers,prb,fleqn,floatfix]{revtex4}

\usepackage{graphicx}
\usepackage{dcolumn}
\usepackage{bm}
\usepackage{amsmath, amssymb}
\usepackage{color}
\begin{document}

\title{The hole Fermi surface in Bi$_{2}$Se$_{3}$ probed by quantum oscillations}

\author{B. A. Piot$^{1}$, W. Desrat $^{2}$, D.K. Maude$^{3}$, M. Orlita$^{1}$, M.
Potemski$^{1}$, G. Martinez$^{1}$}

\affiliation{$^{1}$ Laboratoire National des Champs Magn\'etiques
Intenses, LNCMI-CNRS-UGA-UPS-INSA-EMFL, F-38042 Grenoble, France}

\affiliation{$^{2}$Universit\'{e} Montpellier 2 and CNRS,
Laboratoire Charles Coulomb (L2C), UMR 5221 CNRS-Universit\'e de
Montpellier, F-34095 Montpellier, F-France}

\affiliation{$^{3}$ Laboratoire National des Champs Magn\'etiques
Intenses, LNCMI-CNRS-UGA-UPS-INSA-EMFL, F-31400 Toulouse, France}

\author{Y.S. Hor$^{4}$}

\affiliation{$^{4}$ Department of Physics, Missouri University of
Science and Technology, Rolla, MO 65409, USA}

\date{\today }

\begin{abstract}
Transport and torque magnetometry measurements are performed at
high magnetic fields and low temperatures in a series of p-type
(Ca-doped) Bi$_{2}$Se$_{3}$ crystals. The angular dependence of
the Shubnikov-de Haas and de Haas-van Alphen quantum oscillations
enables us to determine the Fermi surface of the bulk valence band
states as a function of the carrier density. At low density, the
angular dependence exhibits a downturn in the oscillations
frequency between $0^\circ$ and $90^\circ$, reflecting a
bag-shaped hole Fermi surface. The detection of a single frequency
for all tilt angles rules out the existence of a Fermi surface
with different extremal cross-sections down to $24$~meV. There is
therefore no signature of a camel-back in the valence band of our
bulk samples, in accordance with the direct band gap predicted by
$GW$ calculations.
\end{abstract}

\pacs{71.18.+y,71.20.Nr,72.20.My}

%$^{*}$ benjamin.piot@lncmi.cnrs.fr

\maketitle

\section{Introduction}

Bi$_{2}$Se$_{3}$ is a narrow gap layered semiconductor which
together with Bi$_{2}$Te$_{3}$ have been studied for decades for their thermo-electric
 properties.\cite{Mishra1997} The interest in this class of materials
has recently surged because of the prediction \cite{Zhangtheo2009}
and observation \cite{XiaDiraccone2009,ChenBi2Te32009} of a unique
type of charge carriers existing at their surface, the so-called
``helical Dirac fermions'', which behave as massless relativistic
particles with a spin locked to their translational momentum.
Bi$_{2}$Se$_{3}$ therefore now belongs to the 3D topological
insulators family characterized by a bulk gap coexisting with 2D
conducting surface states. As a matter of fact, the existence of
gapless states at the boundary of the material is related to a
well defined change in the \textit{bulk} band structure.  In
Bi$_{2}$Se$_{3}$, this originates from a parity inversion of the
valence and conduction bands at the $\Gamma$ point of the
Brillouin zone in the presence of a large spin orbit coupling.
\cite{Zhangtheo2009,KaneZ2invPRB2007} The linear-in-momentum
dispersion relation which characterizes the 2D surface states thus
emerges from the general Hamiltonian of massive Dirac fermions,
\cite{Zhangtheo2009} theoretically expected to describe the bulk
states in Bi$_{2}$Se$_{3}$.

Pioneering experimental studies of the bulk conduction band at low
energy \cite{KohlerCB1973} have reported an ellipsoidal electron
Fermi surface, which was described within a simple model of
massive carriers with a parabolic (non-parabolic) dispersion in
the k$_{\bot}$ (k$_{\|}$) direction, where k$_{\bot}$ (k$_{\|}$)
is the momentum in the direction perpendicular (parallel) to the
$c$-axis of the crystal. This is accompanied by an increasing
anisotropy of the Fermi surface observed as the Fermi level
increases in the conduction band. More recent transport,
\cite{Eto2010,Fauque2013,Mukhopadhyay2015} NMR,
\cite{Mukhopadhyay2015} and magneto-optics \cite{orlita2015}
measurements have confirmed the original parameters
phenomenologically describing the bulk conduction band, and in
some cases \cite{orlita2015} connected them to the 3D Dirac
Hamiltonian for massive fermions applied to topological
insulators.

However, experimental studies of the \textit{valence band} bulk
Fermi surface are to our knowledge scarce. The principal reason is
that as-grown Bi$_{2}$Se$_{3}$ is electron-doped due to the
presence of Se vacancies. The discovery of 2D surface states has
nevertheless triggered large efforts to reach the topological
insulator regime, where the Fermi level lies in the band gap of
the bulk band structure. For instance, substituting trace amounts
of Ca$^{2+}$ for Bi$^{3+}$ in as-grown Bi$_{2}$Se$_{3}$ can lower
the Fermi energy of the native n-type crystals. Above a certain
value of Ca-doping $\delta$, the electrical conduction in
Bi$_{2-\delta}$Ca$_{\delta}$Se$_{3}$ is supported by hole carriers
rather than electrons. \cite{Horptype2009,Hsiehtunable2009}
Further doping brings the Fermi level deep in the previously
inaccessible valence band. Very recently, Shubnikov-de Haas (SdH)
measurements have been reported \cite{Gao2014} in p-type
Bi$_{2}$Se$_{3}$ samples with hole concentrations estimated
between $5.7\times10^{18}$ cm$^{-3}$ and $1.6\times10^{19}$
cm$^{-3}$. A bag-like closed Fermi surface was observed at low
concentration, with the suggestion of open tubes appearing in the
Fermi surface at high carrier density. In spite of these first
experimental advances, and several theoretical works,
\cite{Mishra1997,Larson2002,Zhangtheo2009,Parker2011} the
low-energy details of the valence band are still not unambiguously
determined. In particular, a local minimum was suggested to form
at the $\Gamma$ point as a consequence of the spin-orbit coupling.
While a camel-back structure is observed in the valence band near
the surface, \cite{Hsiehtunable2009} it is absent in some bulk
measurements. \cite{orlita2015} More recent $GW$ calculations
\cite{Yazyev2012,Aguilera2013} show that the electron-electron
interactions reduce the band gap at the $\Gamma$ point and wash
out the camel-back structure. This issue brings further motivation
to experimentally investigate the bulk valence band, in particular
close to the $\Gamma$ point.

In this article, we present a doping dependent study of the bulk
valence band Fermi surface in Bi$_{2}$Se$_{3}$ in a previously
unexplored low energy range. High quality calcium-doped
Bi$_{2}$Se$_{3}$ crystals are studied by magneto-transport and
torque magnetometry at low temperatures and under magnetic fields
up to 30 T. A high resolution angular dependence of the quantum
oscillations (both Shubnikov-de Haas (SdH) and de Haas-van Alphen
(dHvA)) enables us to map out the Fermi surface of the bulk
valence band states, in the energy range $E_{F}\sim$ 20-60 meV. At
low Fermi energies, a downturn is observed in the angular
dependence of the oscillations frequency between $0^\circ$ and
$90^\circ$, demonstrating a bag-shaped closed Fermi surface.
Importantly, a single frequency dominates the FFT spectra
regardless of the magnetic field orientation, showing that no
camel-back structure is observed for energies down to $\sim$ 24
meV. The existence of a camel-back structure for lower energies is
hardly probable in respect to the experimental $E(k)$ dependence,
which points to a direct band gap. The Fermi surface anisotropy
increases rapidly as the Fermi level goes higher in the valence
band, and pipe-like structures previously reported at high energy
are confirmed and attributed to trigonal warping. The apparent
hole effective mass, defined in the parabolic band approximation,
is obtained by temperature-dependent studies for $B\parallel
c$-axis, and lies in the $0.245\pm 0.015\,m_0$ range for
$E_F\sim23-45$~meV. High magnetic fields measurement in the lowest
density samples enable us to approach the quantum limit for holes
which is finally discussed.

\section{Quantum oscillations in p-type $Bi_{2}Se_{3}$}

\subsection{Experimental details}

The Bi$_{2-\delta}$Ca$_{\delta}$Se$_{3}$ samples studied here were
grown via a process of two-step melting described in Ref.
\onlinecite{Horptype2009}. By adding Ca, a transition to a p-type
behavior is observed for x$ > 0.012$, \cite{WangAPLptype2010}
which is the regime our study focuses on. The samples presented
here are referred to as B2, B3, B6, E2, and E1, and their main
characteristics are summarized in Table \ref{tab1}.
\cite{SI.transport}
\begin{table}[h]
\begin{center}
\begin{tabular}{cccccc}
\hline\hline &Ca doping ($\delta$)& Expt & F (T) & $m_{h}$ ($m_0$) & $E_F$ (meV)\\
\hline

B2 & 0.015 & SdH & 48.7 & 0.238(0.01) & 23.7 \\
B3 & 0.015 & SdH  & 55.8  & - & 27.15 \\
B6 & 0.015 & dHvA  & 87.8 & - & - \\
B1 & 0.015 & SdH  & 96 & 0.248(0.01) & 44.8 \\
D1 & 0.025 & SdH & 97 & 0.249(0.01) & 45.1 \\
E2 & 0.03 & dHvA &  100.8 & - & 46.9\\
E1 & 0.03 & dHvA & 126 & - & 58.6 \\
\end{tabular}
\end{center}
\caption{Parameters of the Bi$_2$Se$_3$ samples. Calcium doping
level $\delta$ defined by Bi$_{2-\delta}$Ca$_{\delta}$Se$_{3}$,
experimental technique used, quantum oscillation main frequency,
apparent valence band effective mass (m$_{h}$) and Fermi energy in
the parabolic band approximation.}\label{tab1}
\end{table}
Magneto-transport experiments were conducted on $\mu$m-thick
slices on which silver paste contacts were deposited in a Hall
bar-like configuration. Measurements were performed using a
standard low frequency lock-in technique in a variable temperature
insert for temperatures ranging from $1.2$~K to $40$~K, up to
magnetic fields of 30 T produced by a 20 MW resistive magnet. The
data were initially symmetrized by changing the polarity of the
magnetic field to check that contact misalignment had a negligible
impact on the analysis (notably the oscillation frequency). For
dHvA torque measurements, samples of thickness varying between 40
and 230 $\mu$m were mounted on a CuBe cantilever which forms the
mobile plate of a capacitive torque meter. The torque signal was
measured with a lock-in amplifier and a capacitance bridge using
conventional phase sensitive detection at $5.3$~kHz. The
experiment was performed using a $16$~T superconducting magnet and
a dilution fridge, equipped with an \emph{in-situ} rotation stage.
The torque measurements, which typically probe a larger number of
particles, were employed to further confirm the domination of the
bulk states in the present study.

\subsection{Magnetic field dependence}\label{Bdep}

\begin{figure}[h]
\begin{center}
\includegraphics[width= 8cm]{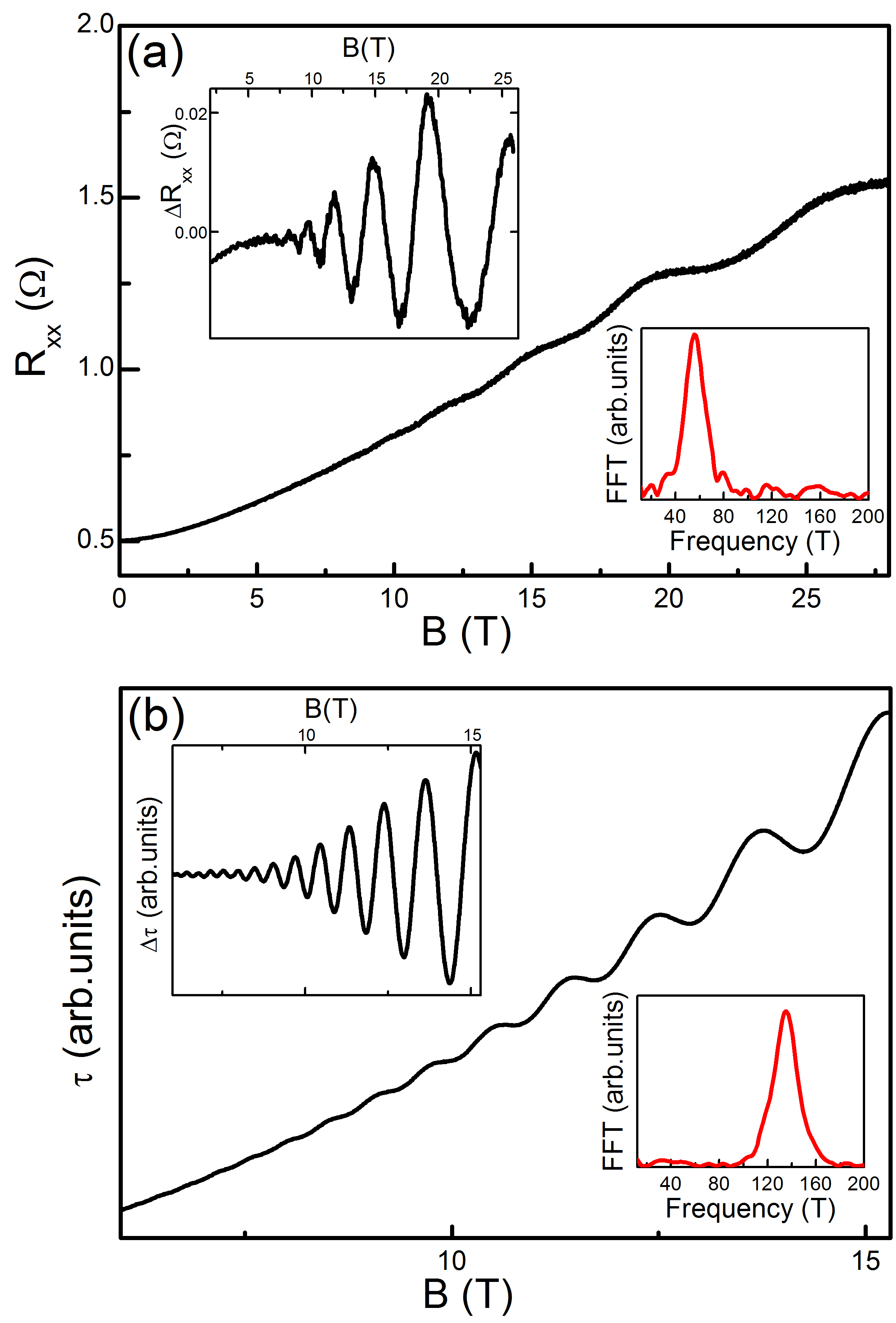}
\end{center}
\caption{(color online) (a) Longitudinal resistance $R_{xx}$
versus total magnetic field in perpendicular configuration for a
low-doped p-type Bi$_2$Se$_3$ sample (B3) at $T=1.2$~K. Upper left
inset: oscillatory resistance $\Delta R_{xx}$. Lower right inset:
fast Fourier transform of $\Delta R_{xx}(1/B)$. (b) Torque $\tau$
and oscillatory torque $\Delta\tau$ (lower right inset) versus
total magnetic field for $\theta=16^{\circ}$ in a highly-doped
p-type Bi$_2$Se$_3$ sample (E1). (Upper left inset) Fourier
transform of $\tau(1/B\cos\theta)$.}\label{Fig1}
\end{figure}

Figure \ref{Fig1} shows representative quantum oscillations in the
valence band for the two extreme carrier densities studied.
In Fig. \ref{Fig1}(a), we report on magneto-transport (SdH) data
on sample B3, one of our lowest density samples. Superimposed on a
large monotonic background, oscillations of the resistivity
reflect the oscillatory density of states of the system. The data
in the upper left inset are obtained by subtracting a smoothed
(moving window average) data curve. The resulting FFT is shown in
the bottom right inset. It shows a clear single peak at about
$55$~T, consistent with the apparent absence of splitting in the
data.

dHvA oscillations are shown for a sample with a higher carrier
concentration (sample E1) in Fig. \ref{Fig1}(b), where the torque
signal $\tau$ is plotted as a function of the total magnetic field
for a tilt angle $\theta = 16^{\circ}$. Small quantum oscillations
in $\tau$, again on a large monotonous background, are clearly
visible and reflect the oscillatory magnetization of the system.
The dHvA oscillations can be better observed in the oscillatory
torque ($\Delta\tau$), obtained with the same background removal
procedure as for transport. Here also a single frequency is
observed, at a higher value consistent with the higher carrier
concentration of sample E1. We note that the results obtained with
the two different experimental techniques for a given carrier
density are fully consistent.

\subsection{Temperature dependence}\label{Tdep}

\begin{figure}[h]
\begin{center}
\includegraphics[width= 8cm]{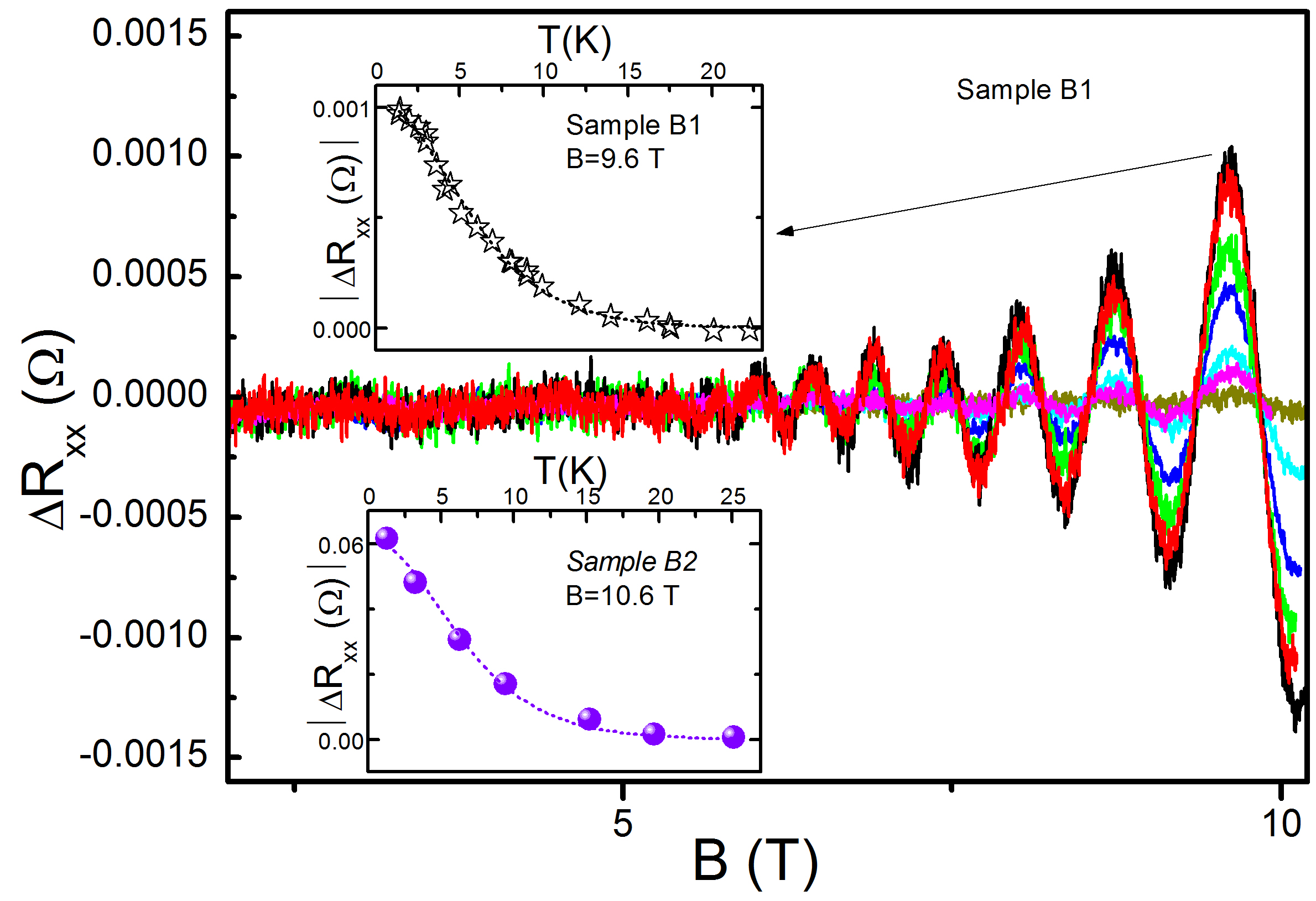}
\end{center}
\caption{(color online) Oscillatory magnetoresistance $\Delta
R_{xx}$ versus total magnetic field parallel to the $c$-axis for
different temperatures (1.4-17.5~K) in sample B1. Upper left
inset: temperature dependence of $\Delta R_{xx}$ at a resistance
extremum at $B=9.6$~T. Lower left inset: temperature
dependence of $\Delta R_{xx}$ at the resistance extremum at
$B=10.6$~T for sample B2.}\label{Fig2}
\end{figure}

The temperature dependence of the SdH oscillations was measured for
samples B1, D1, B2 and B3 between 1.3~K and 40~K for magnetic fields
up to 30~T. In Fig. \ref{Fig2}, we report typical results
obtained on samples B1 and B2 up to 11~T, for temperatures between
1.3~K and 25~K.
In the upper left inset, we plot the temperature dependence of the
oscillation amplitude at the fixed magnetic field $B=9.6$~T. The
same type of data are reported in the lower-left inset for a lower
concentration sample (B2) at a similar magnetic field
($B=10.6$~T). The amplitude of the oscillations has not fully
saturated at $T=1.3$~K, suggesting a higher hole effective mass
compared to the well-documented n-type samples \cite{KohlerCB1973,
Mukhopadhyay2015} in qualitative agreement with recent
magneto-optics studies. \cite{orlita2015} The temperature damping
of the oscillations can be well-described by the standard
Lifshitz-Kosevich formalism \cite{LK56}, valid in this case of 3D
quantum oscillations of moderate amplitude. We extract the energy
gap $\Delta$, from which the apparent effective mass of holes is
defined, m$_{h}$=$(\hbar eB)/ \Delta$ for a given magnetic field
$B$. The values reported in Table \ref{tab1} correspond to an
average value on the lowest magnetic fields data sets exploitable
on wide enough temperature range (typically 8-10~T). The mass was
found to be almost constant over the energy range studied, with a
value of $0.245\pm 0.015\,m_0$. This value is consistent with the
values obtained in a higher (but overlapping) energy range.
\cite{Gao2014} A non-trivial field dependence was observed as the
magnetic field was increased, \cite{SI.massB} but is beyond the
scope of the present paper where we aim at characterizing the
(field-independent) hole Fermi surface. From the frequency $F$ of
the quantum oscillations discussed in the previous section, one
can deduce the extremal Fermi surface cross-section in the
momentum space, $CS=\pi k_{F}^{2}$ (where $k_{F}$ is the Fermi
wave vector), given by $CS=2 \pi e F/\hbar$. In the case of an
energy-independent effective mass in the $B\parallel c$-axis
configuration, the Fermi energy can be written
$E_{F}=(\hbar^{2}CS)/(2 \pi m_{h})$. Using this method, we have
determined the Fermi energies reported in Table \ref{tab1}, which
define the energy range ($E_{F} \sim$ 20 to 60 meV) probed by our
experiment.

\section{Hole Fermi surface}

\subsection{Doping and angular dependences}

\begin{figure}[h]
\begin{center}
\includegraphics[width= 8cm]{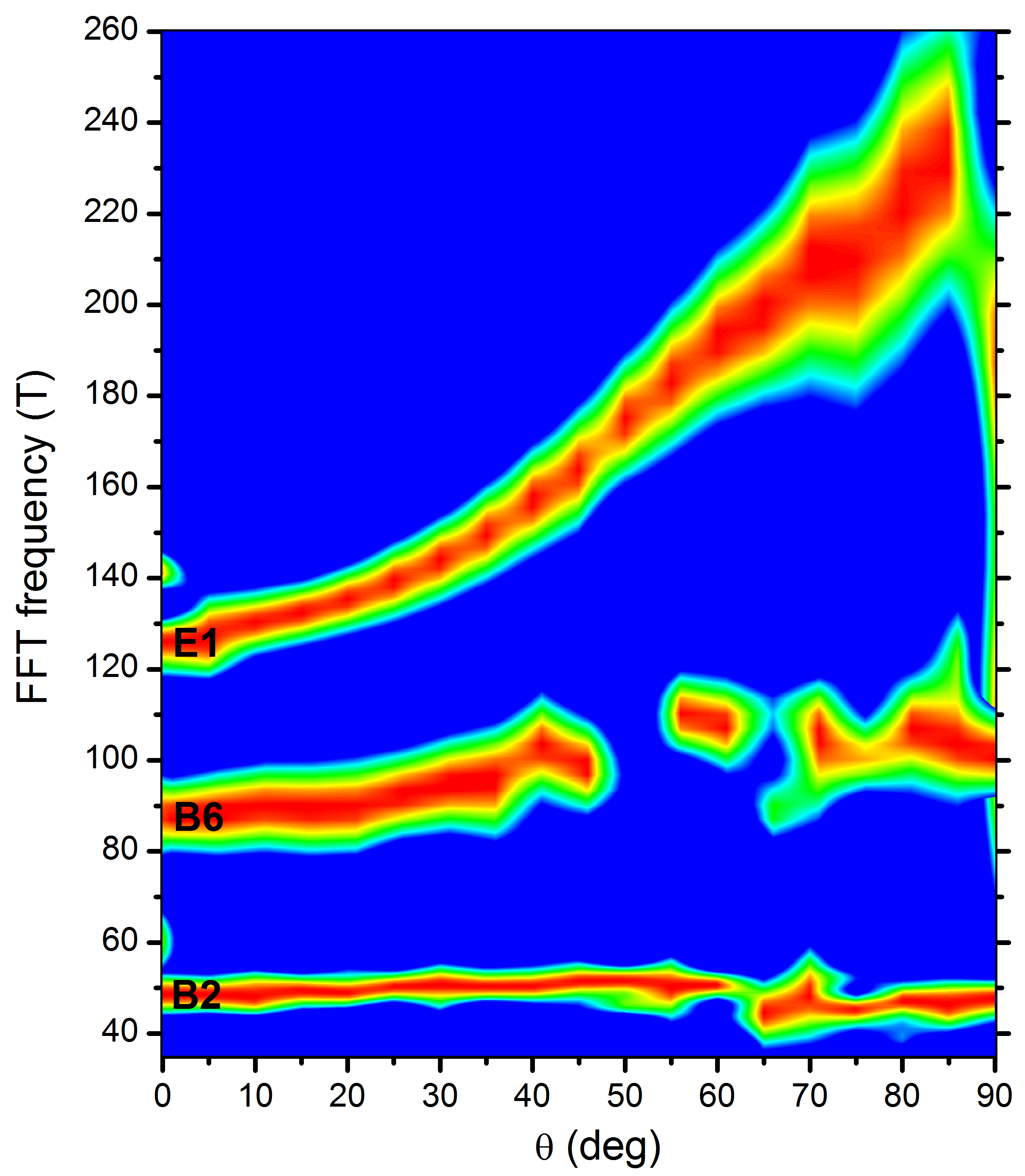}
\end{center}
\caption{(color online) Color plot : \textit{Normalized} amplitude
of the quantum oscillations $1/B$ Fourier transform as a function
of the frequency and the tilt angle $\theta$, for different
samples. At each angle, the FFT signal is normalized to its
maximal value (red color). The blue color corresponds to an
intensity $\leq 70 \%$ of the maximum value.}\label{Fig3}
\end{figure}

In Fig. \ref{Fig3}, we report the angular dependence of the FFT of
the quantum oscillations in a color map for three samples with
different hole concentrations.

The displayed color map is a linear extrapolation of angular
dependent data taken with a 5$^{\circ}$ step, where $\theta$ is
the angle between the sample $c$-axis and the magnetic field. The
FFT signals have been normalized to their maximum value for each
angle to focus \textit{only} on the angular dependence.
\cite{SI.FFTangular} The frequency of the quantum oscillation is
directly related to the cross-section of the Fermi surface in
$k$-space. As the sample is rotated in the magnetic field, the
angular dependence of this cross-section can be traced and related
to the Fermi surface's geometry.

The overall observed non-monotonous behaviour of the frequency as
a function of the tilt angle $\theta$ is characteristic of an
anisotropic Fermi surface. At low energies ($E_F<30$ meV) the
angular dependence is rather mild, suggesting an almost isotropic
Fermi surface. Anisotropy progressively develops with increasing
Fermi energy and, for a frequency of $120$~T at $\theta=0^\circ$,
the $\theta=90^\circ$ frequency approximatively doubles. It should
be noted that a $\cos\theta$ behavior, usually associated with a
2D system, was not observed in any sample even up to very high
magnetic fields. This confirms that bulk states dominate both the
transport and magnetization properties in samples of relatively
large thicknesses. It is worth stressing that, as the system
becomes anisotropic at high doping level, a \textit{complete}
angular dependence ($0^\circ-90^\circ$ at least) is absolutely
required to probe the dimensionality of the system (see the
example of graphite, a very anisotropic 3D system showing a
$\cos\theta$ behavior up to $\sim 70^\circ$ , in Ref.
\onlinecite{GraphitedHvA2012}).

At variance with previous results obtained on n-type
Bi$_{2}$Se$_{3}$ in the low energy region of the conduction band
($E_F<30$ meV), the cross section cannot be reproduced accurately
by assuming a purely ellipsoidal Fermi surface. A perhaps even
more striking difference is the occurrence of a \textit{downturn}
in the angular dependence above a density-dependent angle (e.g.
$\theta \sim 70^{\circ}$ for sample B6). This demonstrates a
``bag-shape'' Fermi surface, where the cross section increases
until it reaches the bag's diagonal axis from where it starts to
decrease with further increasing the angle. A closer look at the
low density data (sample B2) shows the downturn is also present on
the apparently flat angular dependence (see section \ref{theodisc}
for a better representation). The intensity of the FFT signal
(which can not be assessed in Fig. \ref{Fig3} because of the
normalization) can to some extent also be informative. The raw FFT
intensity reflects a severe drop in the oscillation amplitude
where the downturn in frequency appears, for example around
$\theta=60^{\circ}$ for sample B2 and B3. The FFT signal then
clearly reappears from $\theta=75^{\circ}$ to $\theta=90^{\circ}$.
\cite{SI.FFTamplitude} This drop in the amplitude may be
correlated to the downturn in the angular dependence. The FS
strong curvature change in this region probably leads to the loss
of the phase coherence of the oscillations. We note that the
downturn appears at higher angle for higher density. Finally, our
data in the high energy limit (sample E1) are in good quantitative
agreement with the previously reported SdH measurements in high
hole density samples. In particular, oscillations are lost for
specific $\theta$ ranges (see e.g. the sudden drop in frequency
from $\theta=90^{\circ}$), corresponding to open orbits in the
momentum space. We attribute these open orbits to the pipes
emerging in the Fermi surface at high carrier density in the
presence of trigonal warping.\cite{SI.FFThighE}

\subsection{Theoretical model and discussion}\label{theodisc}

\begin{figure}[h]
\begin{center}
\includegraphics[width= \columnwidth]{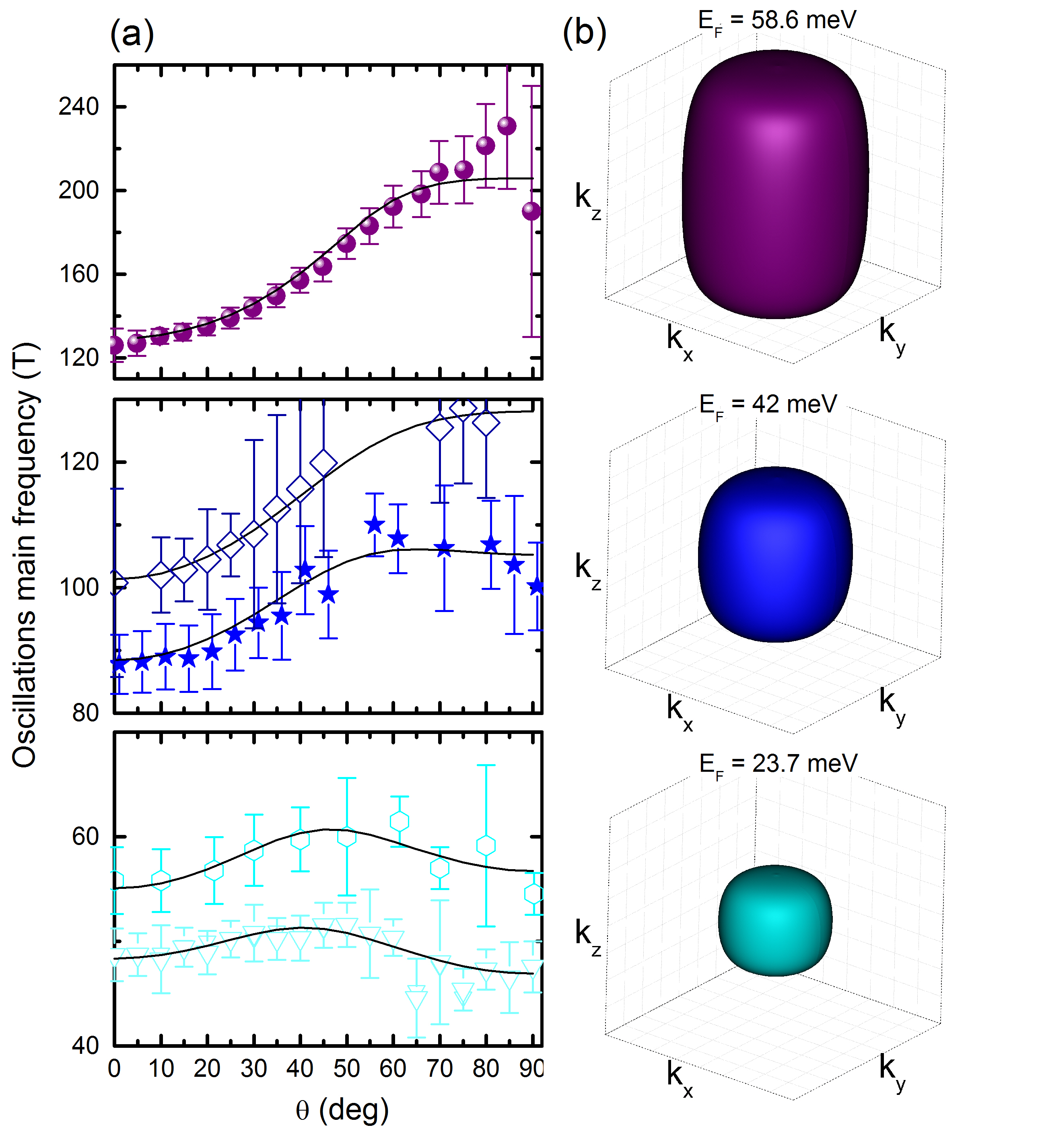}
\end{center}
\caption{(color online) (left panels) Quantum oscillation
frequency as a function of the tilt angle $\theta$ for different
carrier concentrations in the valence band. The error bar is
defined to be the FFT width at 90\% of the maximum signal. From
bottom to top, sample B2 (SdH), B3 (SdH), B6 (dHvA), E2(dHvA) and
E1(dHvA). Theoretical fits obtained by the analytical model of
Eq.\ref{eq1} (see text) (solid lines). (Right panels) Fermi
surfaces computed at $E_F=23.7$, $42$, and $58.6$~meV.}
\label{Fig4}
\end{figure}

In Fig. \ref{Fig4}(a), we report the angular dependence of the
quantum oscillation main frequency (FFT maximum) for different
hole doping levels. These are the same data as the ones in Fig.
\ref{Fig3} with additional samples at intermediate carrier
concentrations.
For each sample we computed the extremal cross-section of the
Fermi surface in $k$-space from the $F(\theta)$ dependence. By
assuming that the Fermi surface is closed and has rotational
symmetry around the $c$-axis, the constant $k_\perp-k_z$ energy
surface, where $k_z$ is parallel to the trigonal axis of the
Brillouin zone ($c$-axis) and $k_\perp=(k_x^2+k_y^2)^{1/2}$ is in
the plane perpendicular to $k_z$, was derived (not shown).
\cite{Mueller66} The $\theta$ angular dependence of the Fermi
radius was then fitted by solving an analytical $E(k)$ dependence
for each experimental Fermi energy reported in Table \ref{tab1}.
The empirical dispersion relation up to the $k^4$ order can be
written as:\cite{kohler75}

\begin{equation}\label{full}
\begin{split}
E(k)=C_1 k_\perp^2+C_2 k_z^2 + C_3 k_\perp^2 k_z^2 + C_4 k_\perp^4
+ C_5 k_z^4  \\
+ C_6 k_x(k_x^2-3k_y^2) + C_7 k_z k_y (k_y^2 -3 k_x^2)
\end{split}
\end{equation}
From symmetry considerations, $E(k)=E(-k)$, which sets $C_6$=0.
Neglecting non-parabolicity effects at low density leads to
$C_4=0$ and $C_5=0$. The last term accounts for the trigonal
warping of the Fermi surface, which we neglect in our first
approach. The simplified expression we used can therefore be
written as:
\begin{equation}\label{eq1}
E(k)=C_1k_\perp^2+C_2k_z^2+C_3k_\perp^2k_z^2
\end{equation}

where the numerical values of the coefficients $C_1$, $C_2$ and
$C_3$ are listed in the supplemental material, section III. The
theoretical $F(\theta)$ dependencies are plotted as solid lines on
top of the experimental points in Fig. \ref{Fig4}(a) and three
computed Fermi surfaces are shown in Fig. \ref{Fig4}(b). At low
energy (sample B2), the downturn of the oscillation frequency is
well reproduced, and clearly results from the bag-shape nature of
the Fermi surface. In this ``squashed'' ellipsoid the in-plane
Fermi wave-vector $k_F$ is longer than along the $c$-axis, leading
to a negative anisotropy factor. \cite{SI.anisotropy} As the
density increases, the shift of the downturn to higher angles is
due to the elongation of the Fermi surface along the vertical
axis. The anisotropy changes its sign, becomes positive, and grows
continuously with $E_F$ up to $1.4$. It is illustrated by the
oblong Fermi surface of sample E2 at $E_F=58.6$~meV.

\begin{figure}[h]
\begin{center}
\includegraphics[width= 0.9\columnwidth]{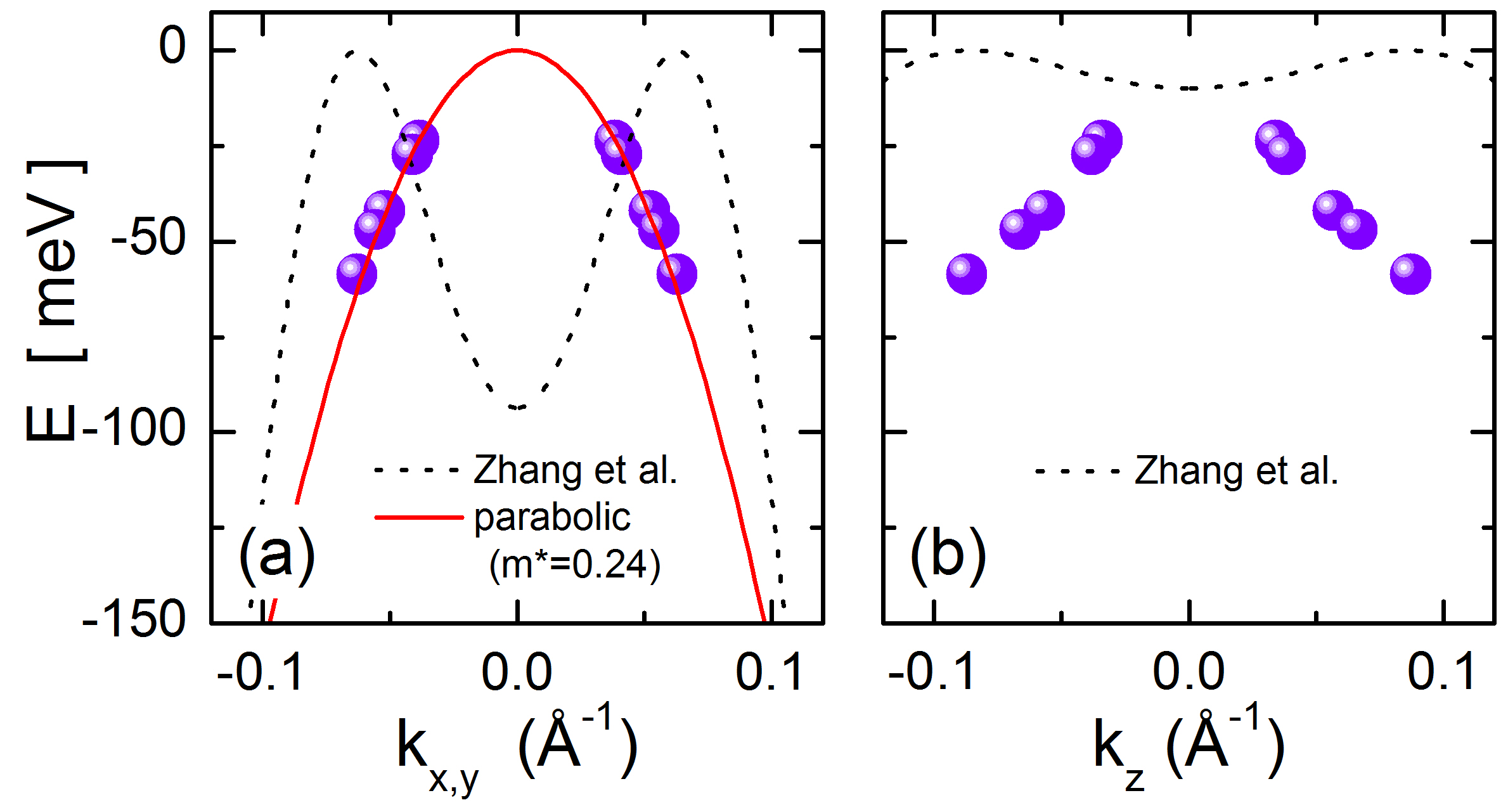}
\end{center}
\caption{(color online)(a) and (b) Valence band along $k_{x,y}$ and $k_z$ respectively. Experimental data (dots), parabolic dispersion (solid lines) and camel-back model (dotted lines).}\label{Fig5}
\end{figure}

We now focus on the low energy limit valence band structure. In
the presence of a strong spin-orbit coupling, the band inversion
is expected to give rise to a camel-back structure, characterized
by a local minimum at $k=0$. This is reported in Figs.
\ref{Fig5}(a) and (b), where the dotted curves are the valence
bands along $k_\perp$ and $k_z$ obtained for the $4\times4$
Hamiltonian of Zhang {\it et al.} \cite{Zhangtheo2009} and its
original parameters. The camel-back is prominent with a band
maximum occurring at finite $k_x\approx0.07\AA^{-1}$ and a $k_x=0$
depth varying in theoretical works from about 80 meV
\cite{Zhangtheo2009} to 140 meV. \cite{Hsiehtunable2009}

If the Fermi energy lies in the camel-back region, there should be
no carriers around $k_{\perp}=0$ and the Fermi surface should
exhibit a ``donut-like'' shape. This should give two frequencies
corresponding the the outer (maximal) and inner (minimal) orbits.
Our observation of a single frequency around $\theta=0^{\circ}$
suggests that there is no ``camel-back'' structure deeper than the
lowest energy investigated ($\sim23.7$ meV). Let us now consider
the case where the residual ``camel back '' depth is lower than
$\sim23$ meV . For our lowest Fermi energy, one would expect a
single oscillation frequency at low angles, splitting into two
frequencies above a certain angle. The low angle frequency would
correspond to an orbit along the horizontal cross section of a
``dumbbell-like'' Fermi surface, and the two frequencies at high
angles would correspond to the two extremal transversal section of
the dumbbell. The size of the frequency splitting at
$\theta=90^{\circ}$ is in this case related to the depth of the
``camel-back'' structure. At variance with these expectations, our
measurements reveals that a \textit{single} oscillation frequency
persists up to $\theta=90^{\circ}$. This shows that no camel back
structure can be observed at all within our experimental
resolution (our FFT full-width-half-maximum at $\theta=90^{\circ}$
is $\Delta F \sim 13$T).

The absence of the camel back structure can more generally be seen
in figure \ref{Fig5}(a), where the experimental points $E_F$ vs
$k_\perp$ are plotted, showing that camel-back structure is
experimentally absent. The $k_\perp^2$ approximation used in Eq.
\ref{eq1} is justified by the good parabolic fit (red solid line)
and the almost energy-independent mass measured experimentally
with the temperature dependent studies $m_{h}=0.24
 m_0$. In Fig.\ref{Fig5}(b), we plot the experimental
dispersion along $k_z$ obtained from our simulations. Importantly,
the $k_z$ dispersion is far from being parabolic, E($k_z$) showing
a slower increase, in agreement with the observation made in
previous ARPES measurements focusing on the $\Gamma-Z$ dispersion
in the Brillouin zone.\cite{Greanya2002} In contrast to the
$k_\perp$ dispersion, higher order terms (appearing in Eq.
\ref{full}) are actually required. These conclusions are similar
to the conduction band results obtained by K\"ohler {\it et al.}
\cite{kohler75}, where the $k_\perp^4$ and $k_z^4$ parameters were
respectively $C_4=0$ and $C_5\neq0$, with $C_5\neq0$ and
energy-dependent. We nevertheless did not consider this term in
the data analysis as it improved the $F(\theta)$ curves plotted in
Fig. \ref{Fig4}(a) only slightly.

Our main observation that no camel-back structure is formed in the
Bi$_2$Se$_3$ valence band is in qualitative agreement with other
experimental work probing the bulk band structure of this
compound. \cite{Nechaev2013,orlita2015} This can be related to the
recent conclusion of $GW$ numerical computations that the band gap
of Bi$_2$Se$_3$ is direct and reduced by the electron-electron
interaction. \cite{Yazyev2012,Aguilera2013}

\section{Hole transport towards the quantum limit}\label{holequantum}

\begin{figure}[!h]
\begin{center}
\includegraphics[width= 8cm]{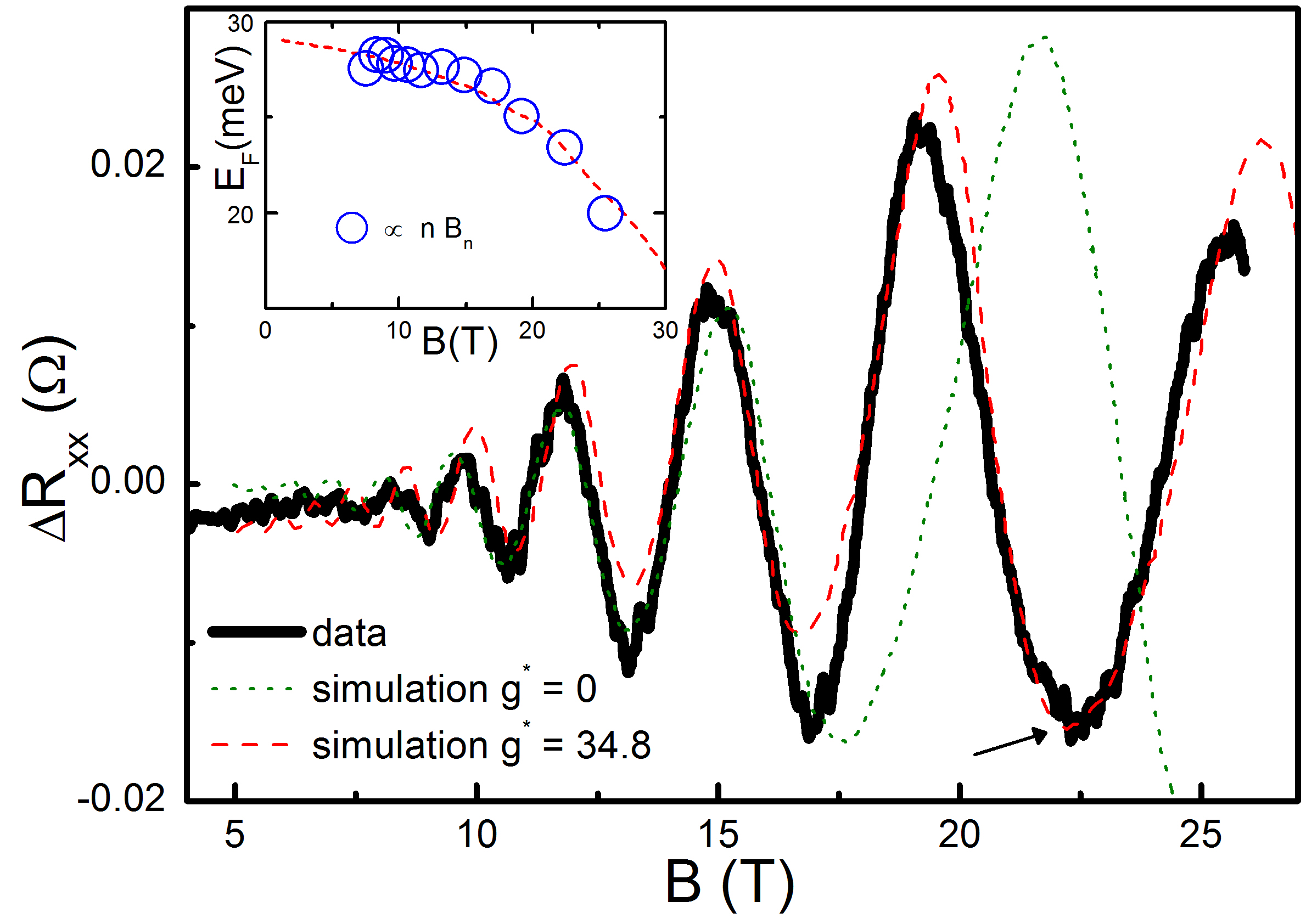}
\end{center}
\caption{(color online) Oscillatory magnetoresistance $\Delta
R_{xx}$ versus total magnetic field at $T=1.2$~K (solid line).
Simulations for the $g^{*}=0$ (dotted line) and $g^{*}=34.8$ cases
(dashed line). The arrow emphasizes the $B=22.5$ T minimum. Inset:
quantity proportional to the $nB_{n}$ value (blue circle), and
simulated Fermi energy as a function of the magnetic field for
$g^{*}=34.8$ (dashed line) (see text)}\label{Fig6}
\end{figure}

For the lowest density samples (B2 and B3), magneto-transport
experiments have been extended to high magnetic fields up to $30$~T.
Typical results are reported in Fig. \ref{Fig6} for sample B3.
Simulations of the magnetoresistance were performed based on the
simple phenomenological approach developed for the conduction
band, \cite{KohlerCB1973, Mukhopadhyay2015} in which one considers
a 3D electron gas with a large (spin-orbit-enhanced) spin gap
$g^{*}_{eff}\mu_{B}B$, where $g^{*}$ is the effective hole
g-factor. As can be seen in Fig.\ref{Fig6} the measured
(background-removed) oscillations exhibit a \textit{minimum} at $B
\sim 22.5$ T, instead of the maximum expected from the $g^{*}=0$
(no spin-splitting) simulation based on the low field oscillation
frequency ($F=55.8$ T). One could tentatively attribute this
additional minimum to a spin-split feature only visible in the
lowest Landau level/highest magnetic fields. The temperature
dependence however shows that the energy gaps associated with the
consecutive minima follow a continuous trend \cite{SI.transport}
and have therefore most likely the same origin. This suggests that
Landau levels of different orbital and spin quantum numbers nearly
coincide in energy, with consequently no visible spin splitting in
the quantum oscillations. A similar scenario is observed in n-type
Bi$_2$Se$_3$ where the spin gap is about twice the value of the
cyclotron gap. \cite{KohlerCB1973,Mukhopadhyay2015} In such cases
the quantum limit is repelled to higher magnetic fields due to a
large field-dependent spin splitting shifting down the Landau
levels energies.

In the present case, when using the effective hole mass
$m_{h}=0.24 m_0$ determined in section \ref{Tdep}, the best
simulation is obtained for a g-factor of $g^{*} \sim 35$ , for
which spin-up and spin-down subbranches of the $N$ and $N+4$
Landau level coincide. As can be seen in Fig.\ref{Fig6}, the
agreement with the data is not perfect (small dephasing can be
seen at low magnetic fields) suggesting that some more complex
field dependencies of the band parameters should be considered for
a better description. The high value of the effective hole
g-factor is nevertheless confirmed by simply analyzing the
magnetoresistance peaks positions. The so-called ``$1/B$ phase
plots'' show a severe deviation from linearity in high magnetic
fields, which can be seen in the inset of Fig.\ref{Fig6} where we
focus on the product $nB_{n}$ where $n$ is an integer and $B_{n}$
the magnetic field value corresponding to the $n^{th}$ SdH
extremum. SdH oscillations extrema occur when the ratio between
the Fermi energy and the cyclotron gap ($\propto B$) equals $n/2$,
and thus the product $nB_{n}$ is proportional to the Fermi energy
value. The y-axis value of the blue circles in the inset of
Fig.\ref{Fig6} is  proportional to the measured $nB_{n}$ value and
show that the Fermi energy is reduced in high magnetic fields.
This deviation is consistent with the Fermi energy drop expected
in the presence of a high Zeeman energy, which is simulated by our
model (dashed line). These observations stress, again, how crucial
the contribution of the Zeeman energy is when conducting a phase
oscillation analysis in such materials, in particular when
searching for the (phase) signature of 2D surface carriers. Our
results suggest that the effective g-factors could be similar in
the bulk conduction and valence bands, which should be confirmed
by a more direct measurement in p-type samples.

In conclusion, we have reported an energy-dependent study of the
Bi$_{2}$Se$_{3}$ valence band based on quantum oscillations
measurements in high magnetic fields and low temperatures. At low
energies, a downturn observed in the oscillation frequency angular
dependence reveals a bag-shaped closed hole Fermi surface. The
absence of a spin-orbit-coupling induced ``camel back'' structure
in the dispersion relation is further demonstrated. As the Fermi
energy increases, anisotropy rapidly develops in the hole Fermi
surface.

We aknowledge support from EC-EuroMagNetII-228043. The material
synthesis work was supported by the National Science Foundation
(NSF) under grant number DMR-1255607.

\end{document}